\documentclass{Pos}
\input{schulzh-preamble.tex}
\newcommand{\pTmin}{\ensuremath{p_\perp^\text{min}}\xspace}

\title{New developments in event generator tuning techniques}

\ShortTitle{New developments in generator tuning}

\author{Andy Buckley\\
        University of Edinburgh\\
        E-mail: \email{andy.buckley@ed.ac.uk}}

\author{Hendrik Hoeth\\
        IPPP Durham\\
        E-mail: \email{hendrik.hoeth@cern.ch}}

\author{Heiko Lacker\\
        Humboldt-University Berlin\\
        E-mail: \email{lacker@physik.hu-berlin.de}}

\author{James Monk\thanks{Many thanks to James Monk who presented this material in our absence.}\\
        UCL\\
        E-mail: \email{jmonk@hep.ucl.ac.uk}}

\author{Holger Schulz\\
        Humboldt-University Berlin\\
        E-mail: \email{hschulz@physik.hu-berlin.de}}

\author{Jan Eike von Seggern\\
        Humboldt-University Berlin\\
        E-mail: \email{vseggern@physik.hu-berlin.de}}

\abstract{
Data analyses in hadron collider physics depend on background simulations performed by Monte Carlo (MC) event generators. However, calculational limitations and non-perturbative effects require approximate models with adjustable parameters. In fact, we need to simultaneously tune many phenomenological parameters in a high-dimensional parameter-space in order to make the MC generator predictions fit the data. It is desirable to achieve this goal without spending too much time or computing resources iterating parameter settings and comparing the same set of plots over and over again. We present extensions and improvements to the MC tuning system, Professor, which addresses the aforementioned problems by constructing a fast analytic model of a MC generator which can then be easily fitted to data. Using this procedure it is for the first time possible to get a robust estimate of the uncertainty of generator tunings. Furthermore, we can use these uncertainty estimates to study the effect of new (pseudo-) data on the quality of tunings and therefore decide if a measurement is worthwhile in the prospect of generator tuning. The potential of the Professor method outside the MC tuning area is presented as well.
}

\FullConference{13th International Workshop on Advanced Computing and Analysis Techniques in Physics Research, ACAT2010\\
		February 22-27, 2010\\
		Jaipur, India}

\begin{document}

\section{MC tuning with \professor}

The ``\professor'' approach to MC tuning constitutes both a numerical method and
a suite of tools which implement it. Fundamentally, \professor attempts to
parameterise expensive functions -- the bin values in a set of MC observables --
by least-squares fitting of the parameterisation coefficients. The least-squares
minimisation is made more approachable by use of the pseudoinverse method,
implemented via a matrix singular value decomposition. Armed with a fast
analytic model of how every bin of a large set of observables will respond to
variations of the generator parameters, numerical optimisation of the
generator's fit to reference data may be efficiently computed. A detailed
description may be found in reference~\cite{Buckley:2009bj}.

The benefit of this approach is clear for more than 2 parameters: \professor
requires as input the values of observables for a moderately large number of MC
runs distributed suitably in the generator parameter space, each point in the
space perhaps requiring \order{48} CPU hours to complete. A serial optimisation
approach such as Markov chain sampling would hence require thousands of CPU days to
have a chance of converging, if the generator itself is not
batch-parallelised. \professor, given sufficiently large batch computing
resources, can trivially parallelise the generation of the input MC points for
any generator and thereafter complete the parameterisation and fit optimisation
in negligible time, allowing for scaling to higher numbers of tune parameters
than could be attempted by methods which require iteration of the time-limiting
step.

\section{Qualitative error estimation in \professor}

An important feature of the \professor method is that it has always allowed for
\emph{qualitative} assessment of the tune robustness. A per-bin parameterisation
in $p$ parameters will require a minimum number of MC runs, \Nmin{p}, for the
least-squares pseudoinversion to be performed. For robustness it is advisable to
oversample this minimum requirement by a factor \order{3} (or more, especially
for large $p$) such that a tune will in fact use $N \gg \Nmin{p}$ input
runs. Additionally, since the parameterisation and optimisation steps are fast,
we take the opportunity to make many such overconstrained tunes by in fact
sampling an even greater number of runs, $N_\text{sampled} \gg N$. We can then
randomly sample a large number of mostly-independent $N$-run tunes to obtain an
ensemble of reasonable tunes -- again, this step can be trivially
batch-parallelised. The spread of this tune ensemble as projected on each
parameter has been used in several \professor MC tunes as a heuristic for
determining whether a parameter is well or poorly constrained, for detecting
parameterisation problems, and for ensuring that the ``maximum information''
tune is typical of the ensemble. Several other checking methods, such as
eigenvector line scans, are also used to ensure that the details of the tune,
and particularly the generator parameterisation, are reliable.

\begin{figure}[t]
  \centering

  \includegraphics[width=0.325\textwidth]{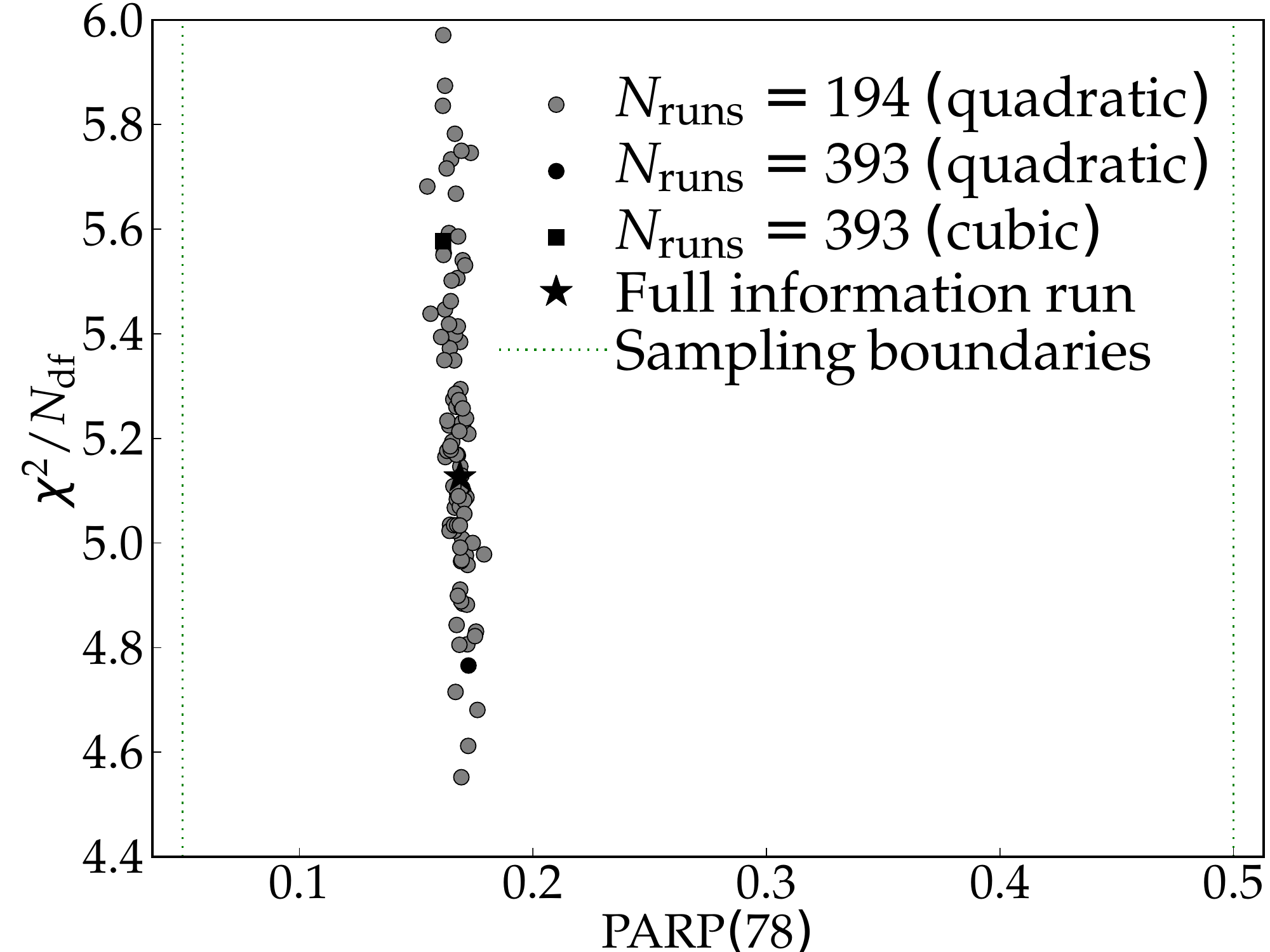}
  \includegraphics[width=0.325\textwidth]{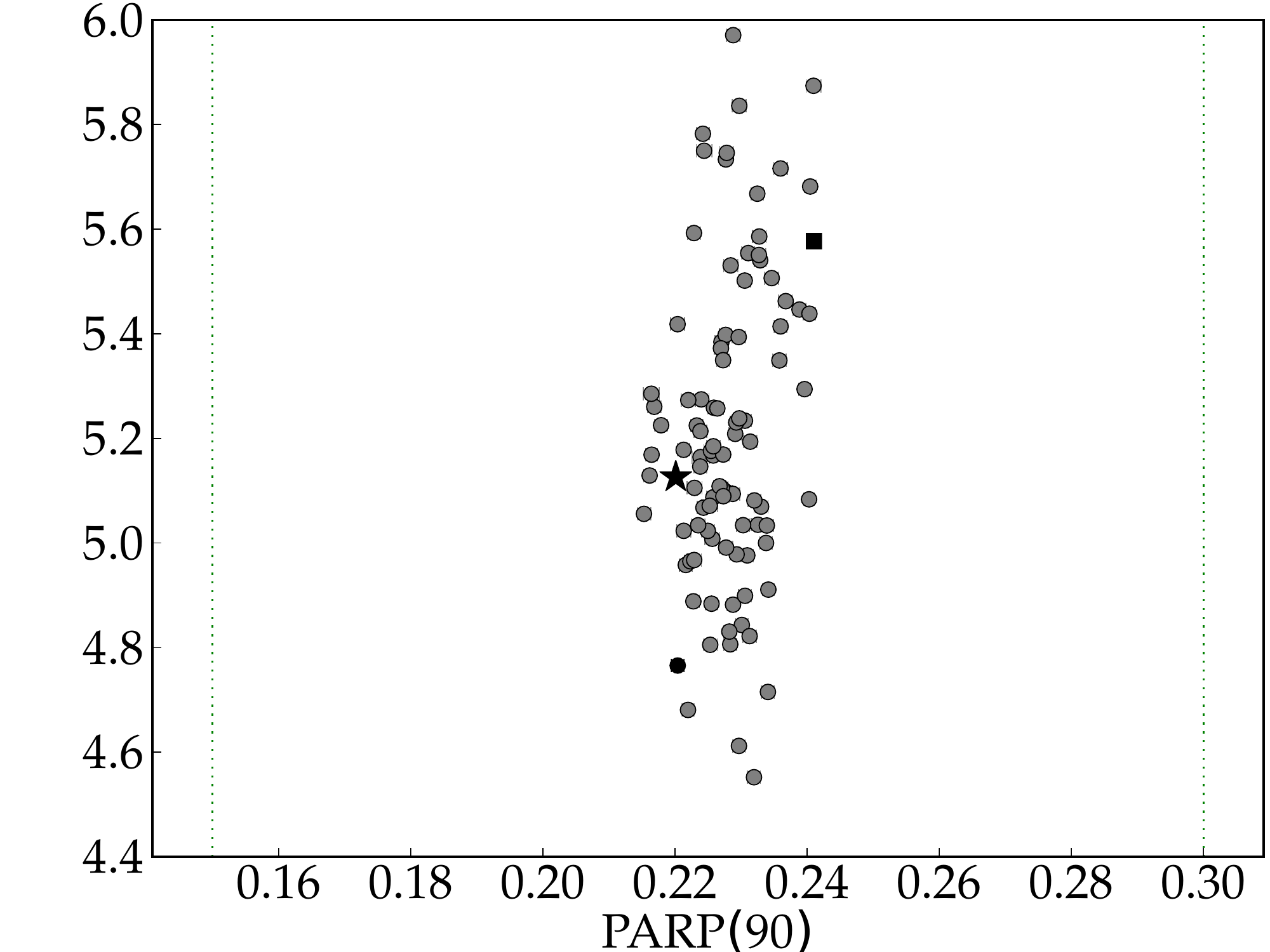}
  \includegraphics[width=0.325\textwidth]{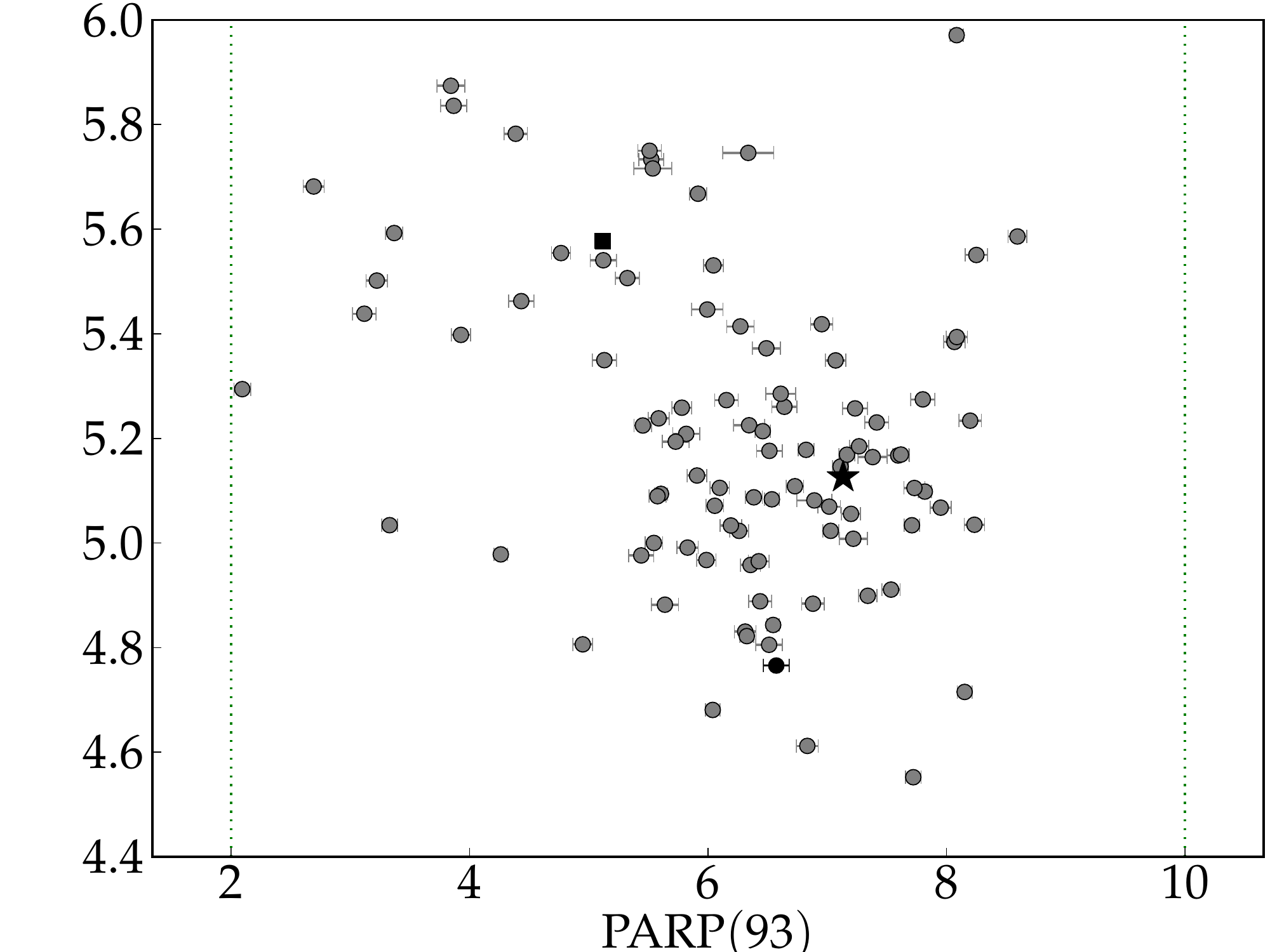}

  \caption{A scatter plot of \chisq vs. parameter value for a set of \professor
    run combinations on three \pythia\!6 parameters. Implicitly, this projection
    of tune parameter vectors on to parameter axes gives a qualitative measure
    of whether or not a parameter is well-constrained: these parameters become
    increasingly ill-defined from left to right.  The different markers
    represent different degrees of oversampling, with the star representing the
    maximum information run --- the points are for the same run combinations in
    all three scatter plots.}
  \label{fig:paramscatter}
\end{figure}
\textbf{}

To make a prediction of an observable for which there is no reference data (this
may be an energy extrapolation, or simply an unmeasured feature at existing
energies), the simplest approach is of course to run the generator with the
obtained tune(s) and compute the observable. Using the many tunes resulting from
the different run combinations will give a spread in the observable prediction,
reflecting one part of tune uncertainty. In practice, since we can build a fast
parameterisation of the generator behaviour (in fact many of them) on the unfitted
observable as part of the main \professor tuning process, this offers a much
faster turnaround than processing another large (perhaps \emph{very} large) set
of generator runs -- with the proviso that the parameterisations are of course
non-exact.

\section{Sources of tune uncertainty}

Before making this method more quantitive, we now consider the various sources
of uncertainty in the procedure outlined so far. This will help us to understand
which sources of uncertainty are computationally controllable and which will
have to, for now, remain more nebulous. These main sources of tuning error are
as follows:
\begin{enumerate}
\item Error on experimental reference data.
\item Statistical error on the MC at the anchor points from which the
  parameterisation is constructed.
\item Systematic limitations of the parameterising function to describe bin
  responses to parameter variations -- pathological MC parameters with
  discontinuous or critical behaviour are particularly hard to generically
  parameterise, since a Jacobian transformation to a suitable meta-parameter is
  not always easily available.
\item Choice of run combinations to make the parameterisation.
\item Goodness of fit definition, including both the type of GoF measure and the
  choice and relative weighting of different data.
\item Reasonable minimiser scatter within the \chisq valley containing the
  optimal tune point for a given parameterisation. Note that this cannot be
  completely disentangled from the role that error sources 1, 2 and 3 play in
  defining the \chisq valley.%
\item Limitations of the parameterisation(s) used to compute the parameterised
  MC value in extrapolated/unfitted observables. Of course, this error doesn't
  exist if the less convenient strategy of re-running the MC generator is used.
\item For completeness, we again highlight the systematic error associated with
  the discrete physics model being tuned: the total error is far from complete
  without considering more than one viable model. Within a given model there are
  also systematic uncertainties, some of which may be quantified,
  e.g. cross-section integration uncertainty and PDFs: the second of these is
  particularly quantifiable due to the existence of error or replica PDF sets,
  themselves expressing reasonable variations in PDF fitting.
\end{enumerate}

Note that, for example, these sources of uncertainty such as the variation
between members of the ensemble of run combinations are not unique errors
introduced by the \professor approach: failing to test different run
combinations does not \emph{eliminate} the error associated with the choice of
anchor runs used! A similar, but unquantifiable, error exists for any form of
manual tuning.

\section{Construction of tune uncertainty confidence belts}

Our approach to quantifying the uncertainties from (combinations of) the sources
listed in the previous section is to construct central confidence belts for
observable bins from the various ensembles of tune results, parameterisations,
etc. we have described. Explicitly, given a large number of reasonable and
equivalent predictions for an observable bin value, we construct a band of given
$P$-value as being the region containing fraction $P$ of predictions, with equal
fractions above and below.

There are many ways to construct such ensembles -- for the purposes of this
study we identify three:

\paragraph{Combination error:} The ensemble from which we construct the
confidence belt is simply the ensemble of predictions from different run
combinations. This will hence represent the variation due to error sources 1, 2,
and 4 -- the other sources of uncertainty exist, but are not quantified by this
approach\footnote{In the results shown here, MC error has not been
  \emph{explicitly} propagated into \professor's fit measure, due to
  extrapolation problems: it enters implicitly via the statistical scatter of MC
  samples. This has been remedied in the latest development versions of
  \professor, which use the median MC error of the sampled anchor points to
  avoid the instabilities.}.%
This approach requires that correlations between different run combinations are
small, which is ensured by the $N_\text{sampled} \gg N$ requirement.
If parameterisation (as opposed to explicit MC runs) is used for the translation
of this tune ensemble into bin value predictions, then source 7 also
applies. This can be included into the band construction by using many
parameterisations, again constructed from run combinations. Different
parameterisations should be used for minimisation and prediction to avoid
reinforcing parameterisation systematics.

\paragraph{Statistical error:} The obvious failure of the combination error
approach is that error source 6 -- the measure of reasonable tune variation
within the \chisq valley -- is left unquantified. Hence it will be no surprise
that our ``statistical'' error band is constructed explicitly from an ensemble
of samples from this valley. This is obtained in the simplest case by only using
the maximum information \professor tune -- that which is constructed from all
the available MC runs. The covariance matrix of the parameters in the vicinity
of the tune point is obtained -- in principle directly from the
parameterisation, in practice from the minimiser -- and used to define a rotated
hyper-Gaussian probability distribution in the parameter space. Sampling
parameter points from this distribution gives us another ensemble of tune points
and, as for the combination error, they can be mapped into observable
predictions either by direct MC runs or by using one or many constructed
parameterisations. In practice, we take advantage of the combined potential of
the hyper-Gaussian sampling and MC parameterisation to build a confidence belt
from \order{10,000} samples. The quantified sources of error are hence 1, 2, 6,
and 7.

These approaches to error band building will be used in the next section to
construct sample error bands for underlying event predictions. Although not
currently implemented in \professor, we also highlight the most complete form of
quantifiable error band within the \professor approach:

\paragraph{Combined error:} This extension is an obvious fusion of the above two
ensemble/band constructions: as in the combination error, we construct an
ensemble of points from run combinations, then for each run-combination point we
construct a statistical ensemble. The combined ensemble of tune ensembles, and a
variety of parameterisations to transform them into predictions will lead to
error bands quantifying error sources 1, 2, 4, 6, and 7.

The remaining error sources are 3, 5 and 8: limitations of the parameterisation,
the observable weights/goodness of fit definition, and -- most importantly --
the uncertainty due to different physics models. These missing systematics
remain qualitative in this scheme, and reliable MC predictions should take care
to include estimates of their influence, albeit in a more ad hoc fashion than
for the more statistically-induced errors.


\section{Results}

We now briefly present results using the first two confidence belt definitions
presented above -- ongoing work is addressing the ``combined'' belt definition
and inclusion of PDF uncertainties.

\begin{SCfigure}[1.8][t]
  \scalebox{.25}{
    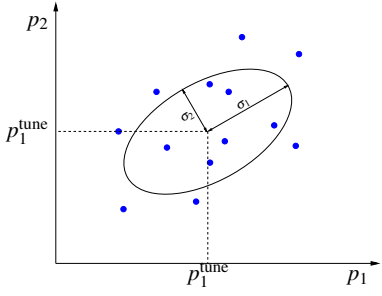
  }
  \hspace{0.5cm}
  \caption{Two dimensional illustration of the parameter point sampling used for
    the statistical uncertainty estimate. We exploit the covariance matrix returned
    by the minimiser for a Gaussian sampling from the corresponding $p$-dimensional
    hyper-ellipsoid. The $\sigma_i$ ($i=1\ldots p$) are the eigenvalues obtained from
    an eigen-decomposition of the covariance matrix.}
  \label{fig:stat-sampling}
\end{SCfigure}

Our exploration is based on tunes of the \jimmy~\cite{jimmy} MC generator, which
simulates multiple parton interactions (MPI) for
\herwigsix~\cite{Corcella:2002jc}, because it has only two relevant parameters,
PTJIM and JMRAD(73)\footnote{We treat the inverse radius-squared of the protons,
  JMRAD(73), to be identical to that of the anti-protons, JMRAD(91), and ignore
  the interplay with ISR parameters in \herwigsix itself.} -- the frugality with
parameters makes \jimmy an ideal ``toy model'' testbed, while remaining
phenomenologically relevant. As a \jimmy-like MPI model is ruled out by Tevatron
data~\cite{Bahr:2008wk}, we fix a dependence of PTJIM on the centre
of mass energy with the same ansatz as used in
\pythiasix~\cite{Sjostrand:2006za}:
\begin{align}
  \label{eq:ptjim}
  \text{PTJIM} =
  \text{PTJIM}_{1800}\cdot\left(\frac{\sqrt{s}}{1800~\text{GeV}}\right)^{\:0.274},
\end{align}
where $\text{PTJIM}_{1800}$ is the value of PTJIM at the reference scale
$\sqrt{s}=1800~\text{GeV}$ and is the \pTmin parameter actually used in the
tuning process. Furthermore, we use the MRST LO* PDF set~\cite{Sherstnev:2007nd} and
use \tevatron data from \cdf~\cite{Affolder:2001xt, Acosta:2004wqa,
  cdf-leadingjet} and \dzero~\cite{Abazov:2004hm} as a tuning reference. A more
complete tune would include the exponent of the \pTmin energy dependence, but
for toy-study purposes we here fix it to a value consistent with known energy
extrapolation fits~\cite{Buckley:2009bj,Skands:2009zm}.

\begin{figure}[t]
  \centering
  \subfigure[Statistical uncertainty, tuning to data only]{\label{fig:statband-Nch}
    \includegraphics[width=0.48\textwidth]{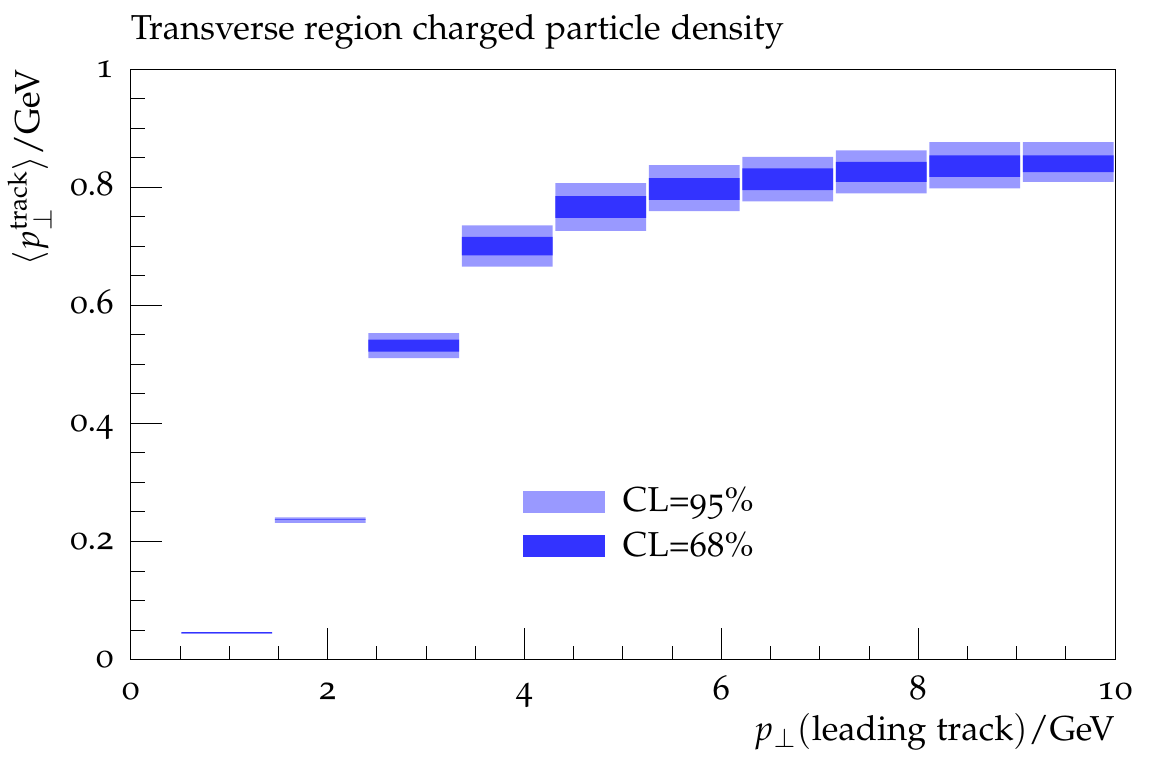}
  }
  \subfigure[Statistical uncertainty, tuning to data and pseudodata]{\label{fig:statband-Nch-pseudo}
    \includegraphics[width=0.48\textwidth]{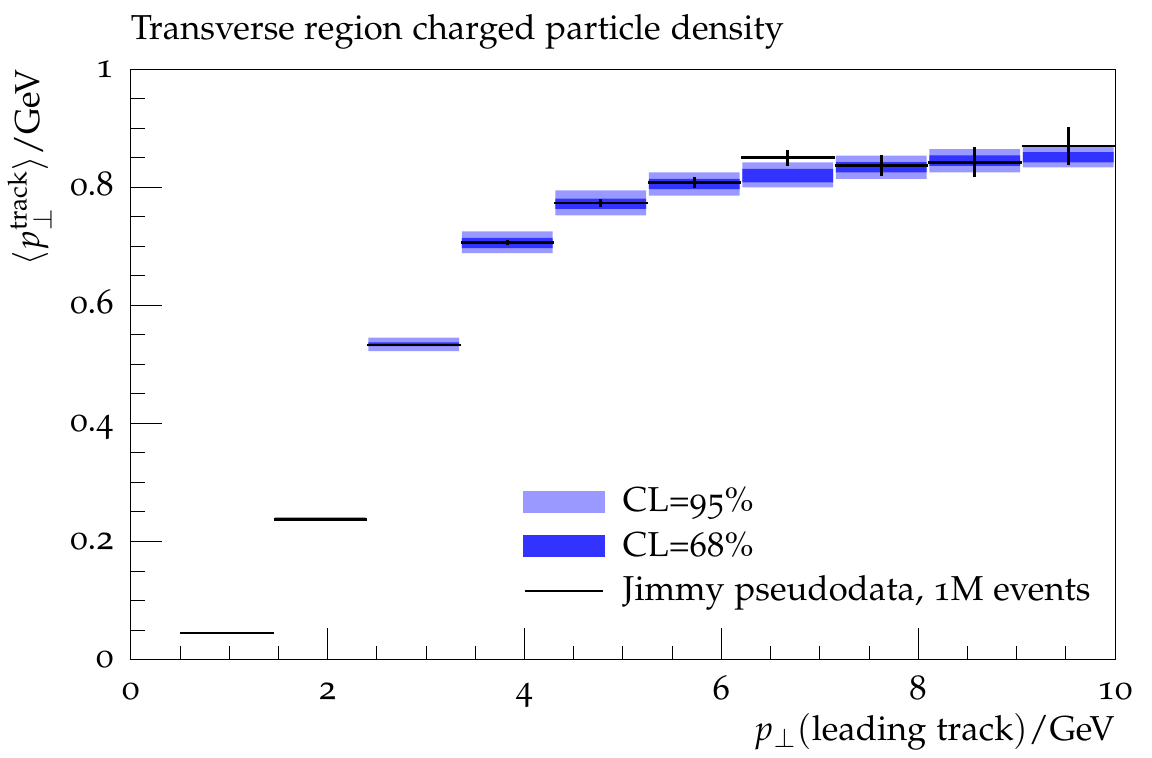}
  }\\
  \subfigure[Combination uncertainty, tuning to data only]{\label{fig:sysband-Nch}
    \includegraphics[width=0.48\textwidth]{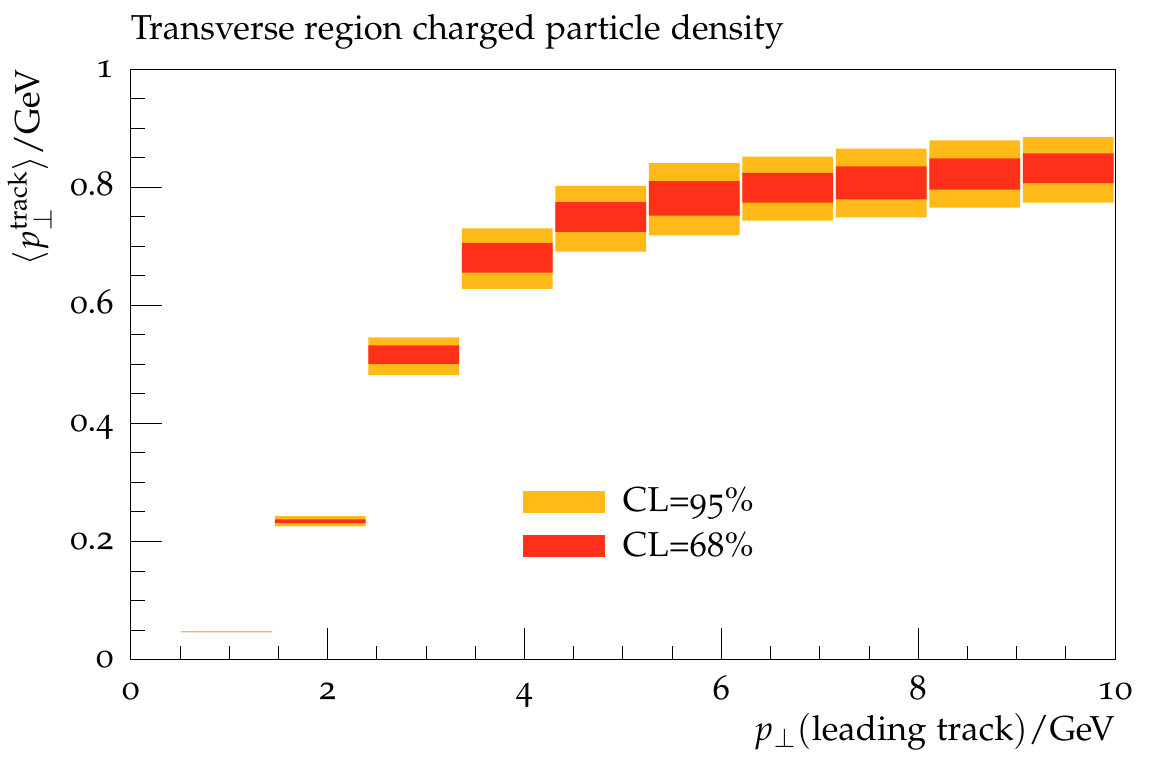}
  }
  \subfigure[Combination uncertainty, tuning to data and pseudodata]{\label{fig:sysband-Nch-pseudo}
    \includegraphics[width=0.48\textwidth]{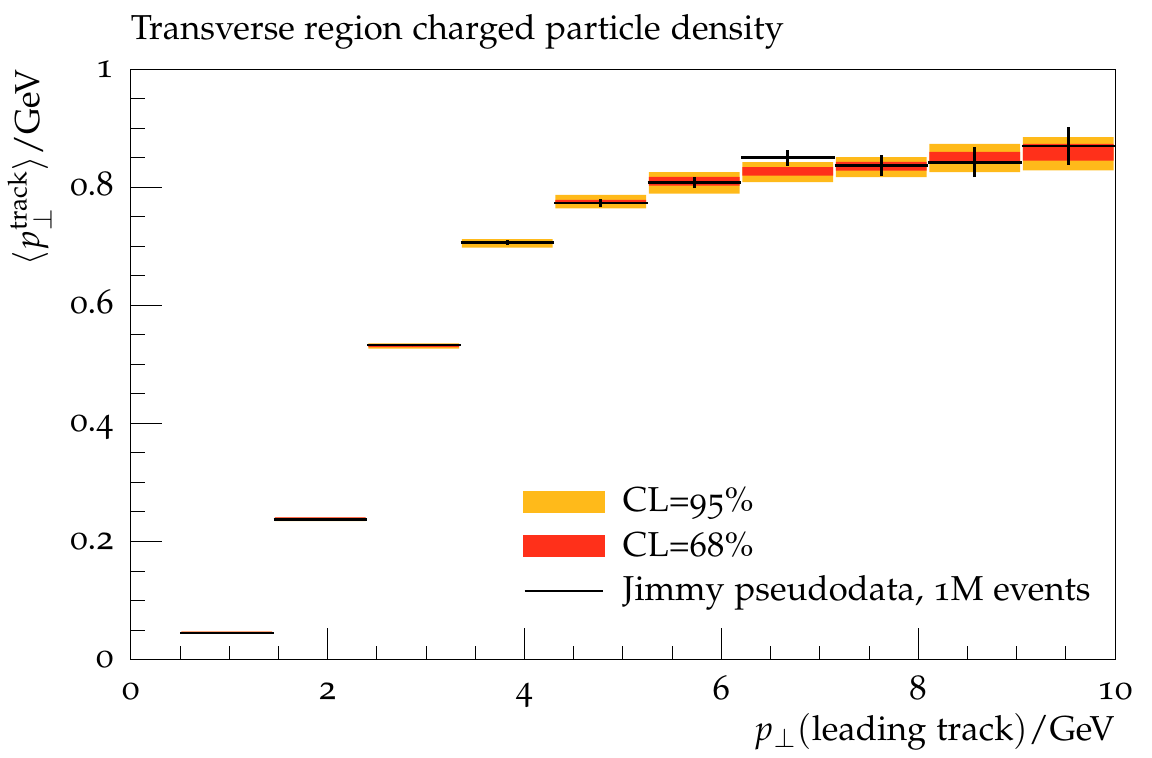}
  }
  \caption{``Statistical'' and ``combination'' error bands for transverse
  $N_\text{ch}$ flow at \unit{7}{\TeV} before and after adding 1M events of
  pseudodata of this observable (black markers) to the tuning. The error bands
  are calculated from the central 95 (68)\% of the binvalues of an ensemble of
  10000 histograms each.}
  \label{fig:statsysbands-Nch}
\end{figure}

In Figures~\ref{fig:statband-Nch} and \ref{fig:sysband-Nch}, the statistical and combination error band
definitions are shown for a \unit{7}{\TeV} underlying event observable --
$N_\text{ch}$ flow transverse to the leading track (track with the largest
transverse momentum in an event), as a function of leading track \pT -- computed
in \rivet~\cite{Waugh:2006ip}, based on the fit to Tevatron reference data. The
error due to variation of run combinations (error source 4) is notably somewhat
larger than the scatter of points in the \chisq valley (error source 6),
indicating that in the \jimmy model the parameters have a strong influence on
this observable, and are hence well-constrained. In
Figures~\ref{fig:statband-Nch-pseudo} and \ref{fig:sysband-Nch-pseudo}, similar
band constructions are shown, but in this case the same \rivet analysis has been
used to simulate the effect of adding 1M events of UE data at \unit{7}{\TeV}
into the fit: the size of both error bands is reduced, as expected.

%


\subsection{Effect of extrapolation}

Finally, we consider systematically how error bands constructed in this way
behave as extrapolations are taken further from the region of constraining
data. In this case, since the computational requirements are significant, we
only consider the ``statistical'' error band.

\begin{figure}[t]
  \begin{center}
    \subfigure[$N_\text{ch}$ vs. \pTlead, plateau]{
      \includegraphics[width=.45\textwidth]{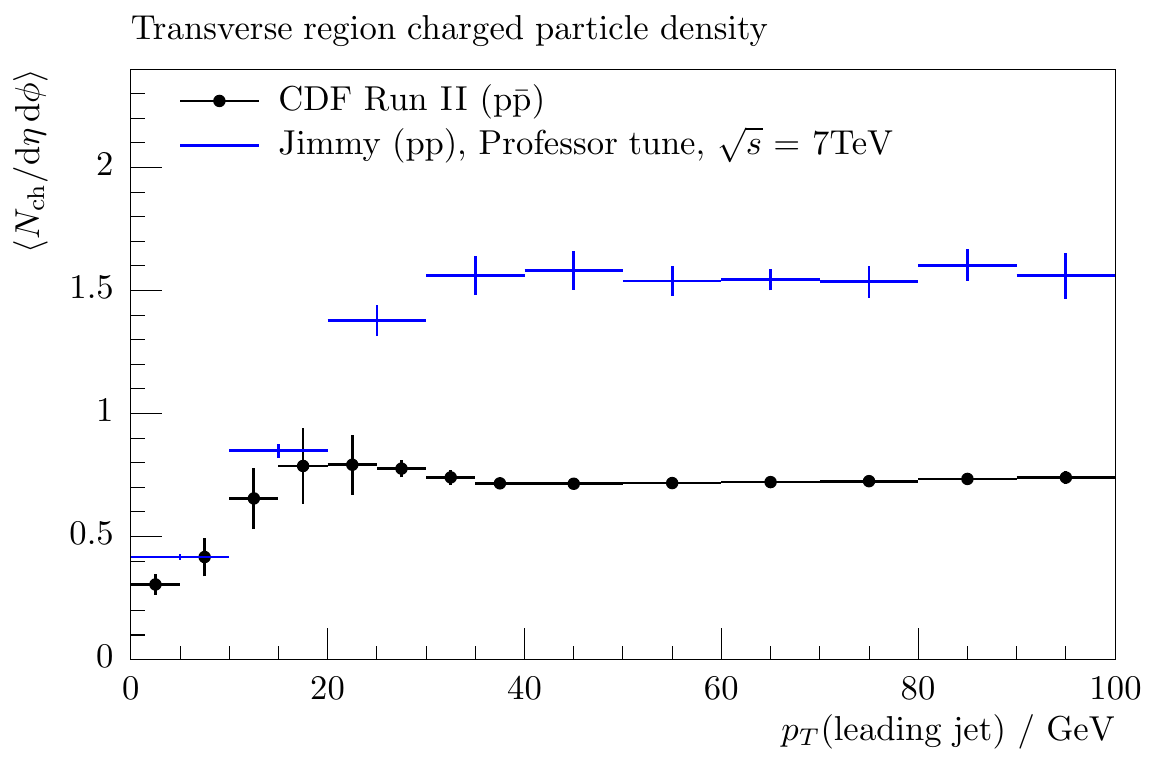}
      \label{fig:transv-sub1}
    }
    \subfigure[$N_\text{ch}$ vs. \pTlead, mean of plateau vs. $\sqrt{s}$]{
      \includegraphics[width=.45\textwidth]{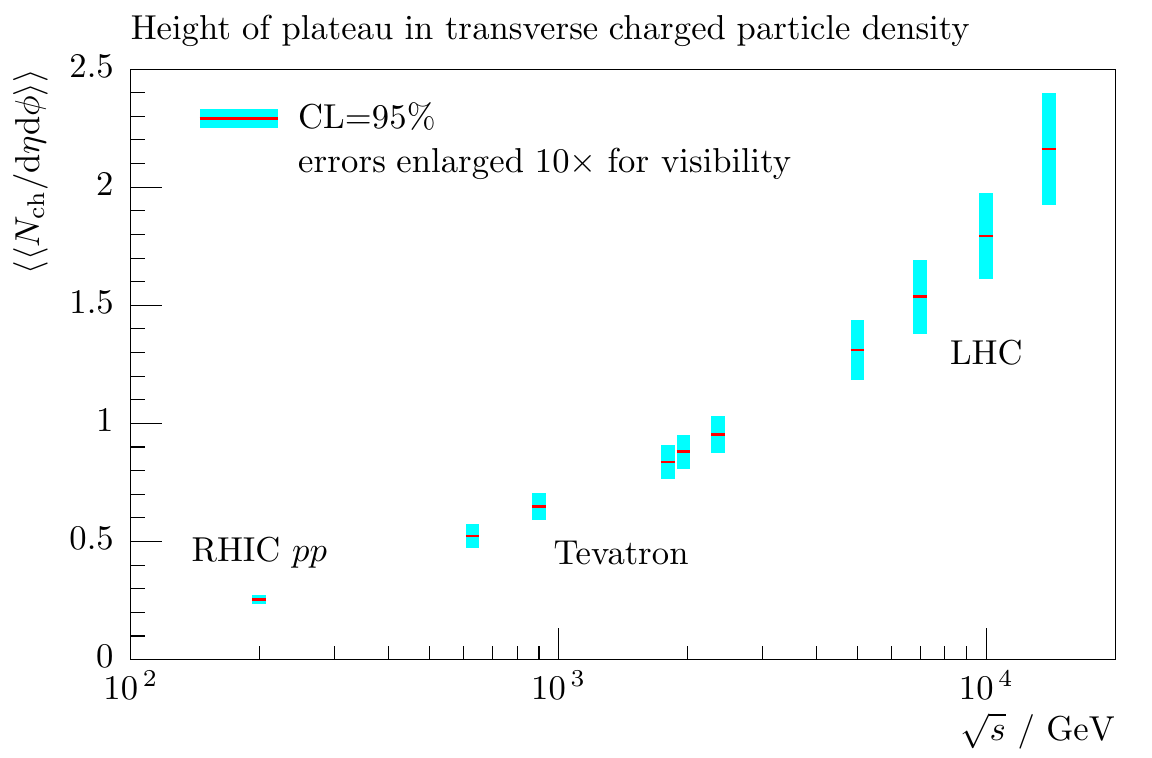}
      \label{fig:transv-sub2}
    }
  \end{center}
    \label{fig:transv}
\end{figure}

We use the transverse region $N_\text{ch}$ density UE observable, evaluated at a range of ten centre
of mass energies, $\sqrt{s}_i$, between 200~\GeV and 14~\TeV. We then apply the
following procedure,
\begin{itemize}
\item Construct the maximum-information parameterisation of the generator
    response for the $N_\text{ch}$ UE observable, $f(\sqrt{s}_i)$
\item Produce an ensemble of 10000 histograms, $H_i$, of the observable shown in
  \FigRef{fig:transv-sub1} (blue line) using the corresponding $f(\sqrt{s}_i)$
  and points sampled using the procedure illustrated in \FigRef{fig:stat-sampling}
\item Calculate the mean height of the $N_\text{ch}$ plateau, $M_i$, for each of
  the $H_i$
\item Construct a $95\%$ central confidence belt, $\mathrm{CL}(\sqrt{s}_i)$ from the $M_i$
\end{itemize}

In \FigRef{fig:transv-sub2} the $\mathrm{CL}(\sqrt{s}_i)$ are drawn. We observe a very
tight confidence belt for the energy region of the \tevatron experiments,
while the confidence belt becomes wider for extrapolation to \lhc energies. The
definitions of the plateau regions used can be found in \TabRef{tab:plateaudefs}.

It is notable that these bands are narrow -- sufficiently so that they
have been visually inflated by a factor of 10 in the figure. While
this reflects good stability in the tuning system, it is probably an
underestimate of the true model uncertainty. A more complete study
will include the exponent of energy extrapolation in
equation~\eqref{eq:ptjim} in the tune, since this may be a dominant
effect in the errors for this particular observable and its inclusion
will give more freedom for various features of the model to balance
against each other: the toy tuning of the energy-dependent \jimmy
model shown here is probably too restrictive to accurately represent
the full range of variation allowed by the model, but serves as an
indication of the extent to which such studies can be systematised.


\begin{table}[tb]
\begin{center}
  \begin{tabular}{lrrrrrrrrrr}
      \toprule
      $\sqrt{s}/\TeV$ & 0.2 & 0.63 & 0.9 & 1.8 & 1.96 & 2.36 & 5.0 & 7.0 & 10.0 & 14.0 \\ 
      \midrule
      \pTleadmin/\GeV & 10  &  30  & 30  & 30  & 40   &  40  &  40 & 40  & 40   & 40 \\   
      \pTleadmax/\GeV & 30  &  70  & 80  & 80  & 90   & 110  &  110 & 120 & 150  & 160 \\
      \bottomrule
  \end{tabular}
  \caption{Definition of plateau regions (\pTlead) used in the extrapolation study.}
  \label{tab:plateaudefs}
\end{center}
\end{table}

\section{Interactive parameterisation explorer}
The \professor package also contains the program \kbd{prof-I} that allows for
interactive exploration of the effect of changing parameter values on the shape
of observables. It makes use of the functionality of the
WXPython\cite{dunn:2010} library to create a graphical user interface (GUI).
The user can choose a Professor interpolation to explore and overlay
experimental and Monte Carlo generator (pseudo-) data for two different
observables.

The heart of this program are simple sliders (one for each parameter in the interpolation)
that can be adjusted freely within certain values of the corresponding parameter.
Exploring the parameter space is done simply by moving the sliders which results in
real-time redrawing of the interpolation's prediction for the two observables studied.

By doing so it one immediately gets a prediction of the generators outcome at a
certain parameter point, a task that by running the generator alone usually takes
many hours or days. This program in principle enables the user to do manual tunings
tool, however, the much greater value of this tool is that it helps to get a
feeling for the effects of shifts in parameter space on certain observables.
Also the decision which observables one should include in a tuning can be helped
since the sensitivity of observables to parameters can be studied with the
\kbd{prof-I} GUI.

\begin{figure}[t]
  \begin{center}
      \includegraphics[width=.95\textwidth]{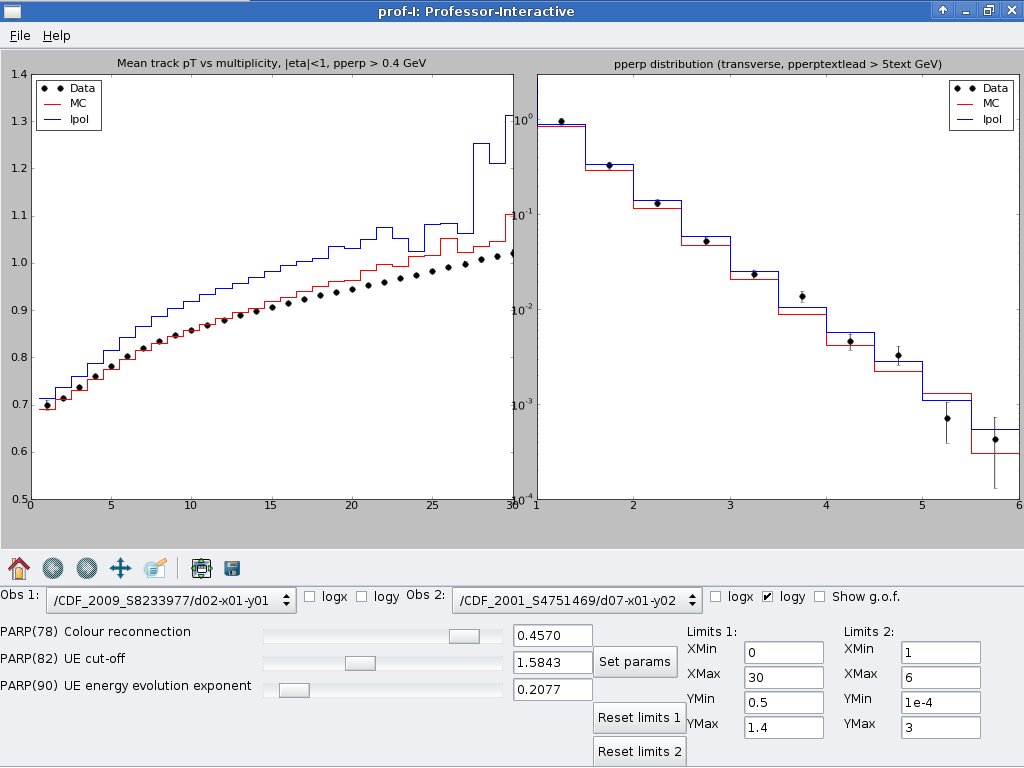}
  \end{center}
  \caption{A screenshot of the interactive parameter explorer of the \professor package, \kbd{prof-I}. The blue line
  represents the interpolation's prediction for points in parameter space that are adjusted by the sliders at the bottom
  of the GUI: moving the sliders results in real-time redrawing.}
    \label{fig:profi}
\end{figure}

\section{Conclusions}

We have catalogued a set of sources of uncertainty which either explicitly or
implicitly contribute to any tune of a MC event generator, and presented a
systematic approach to quantifying many of these uncertainties using the fast MC
parameterisations and natural tune ensembles which arise from the \professor
tuning approach. Example results have been shown, which exhibit some expected
behaviours, such as the shrinking of error bands on adding new reference data in
new areas of parameter space and the blow-up of error bands as predictions
venture further into unconstrained regions.

Several things should be emphasised: first and most important is that this
approach does not catch all sources of error. We have presented results from two
definitions and have proposed a third, more comprehensive measure, but still
variations such as PDF errors and discrete model variations need to be
included. However, with the increased usage of systematic tuning methods,
variations between models in UE observables are not as substantial as once they
were -- statistical errors are a non-negligible factor in assessing the
reliability of phenomenologically-based MC predictions.

The approach taken here has many obvious parallels in the world of PDF errors,
with our approach having a good deal of overlap with the MC replica set approach
of the NNPDF collaboration~\cite{Ball:2008by} as contrasted with the eigenset
approach of the CTEQ and MRST/MSTW collaborations. While replica sets have the
advantage of a more direct statistical uncertainty interpretation (although we
do not have the option of the parameterisation-freedom exibited by the NNPDF use
of neural networks), there is the pragmatic issue that \order{10} representative
error tunes would be more usable than \order{10000} equivalent tunes. Work to
add ``eigen-tune'' calculation to \professor has begun, and we hope to soon
provide statistically-driven rivals to the Perugia-soft/hard tunes for
systematic sensitivity studies.



\section*{Acknowledgements}
We would like to thank James Monk for presenting this material in our absence.
Our particular thanks goes to Luigi~del~Debbio and Richard~Ball for discussions
about statistical coverage and MC error estimation. The Professor collaboration
acknowledges support from the EU MCnet Marie Curie Research Training Network
(funded under Framework Programme 6 contract MRTN-CT-2006-035606) for financial
support and for many useful discussions and collaborations with its members.
A.~Buckley additionally acknowledges support from the Scottish Universities
Physics Alliance (SUPA); H.~Schulz acknowledges the support of the German
Research Foundation (DFG).

{\raggedright
  \bibliography{lh-tune-errors}
}

\end{document}